\documentclass[11pt,a4paper]{JHEP3}
 \usepackage{graphicx}
 \usepackage{amsmath}

\def \be {\begin{equation} }
\def \ee {\end{equation}}

\def \bem {\begin{multline}}
\def \eem {\end{multline}}

\def \bes {\begin{subequations} }
\def \ees {\end{subequations}}

\def \k {\kappa}

\def \l {\lambda}
\def \m {\mu}
\def \n {\nu}

\def \t {\tau}

\def\pa{\partial}

\def\nn {\nonumber}
\def\qquad{\,\, ,\,\,\,}

\title{\LARGE\bf Inverse Magnetic Catalysis from improved Holographic QCD in the Veneziano limit}

\author{Umut G\"ursoy,$^1$\, Ioannis~Iatrakis,$^1$\, Matti J\"arvinen,$^2$\,
and Govert Nijs$^1$
\\
$^1$Institute for Theoretical Physics and Center for Extreme Matter and Emergent Phenomena, Utrecht University, Leuvenlaan 4, 3584 CE Utrecht, The Netherlands
\\
$^2$Laboratoire de Physique Th\'eorique de l' \'Ecole Normale Sup\'erieure \&  Institut de Physique Th\'eorique Philippe Meyer, PSL Research University,  CNRS,  Sorbonne Universit\'es, UPMC Univ.\,Paris 06, 24 rue Lhomond, 75231 Paris Cedex 05, France}

\abstract{
We study the dependence of the chiral condensate on external magnetic field in the context of holographic QCD at large number of flavors. We consider a holographic QCD model where the flavor degrees of freedom fully backreact on the color dynamics. Perturbative QCD calculations have shown that $B$ acts constructively on the chiral condensate, a phenomenon called ``magnetic catalysis''. In contrast, recent lattice calculations show that, depending on the number of flavors and temperature, the magnetic field may also act destructively, which is called ``inverse magnetic catalysis''. Here we show that the holographic theory is capable of both behaviors depending on the choice of parameters. For reasonable choice of the potentials entering the model we find qualitative agreement with the lattice expectations. Our results provide insight for the physical reasons behind the inverse magnetic catalysis. In particular, we argue that the backreaction of the flavors to the background geometry decatalyzes the condensate.   }


\preprint{}

\begin{document}

\newpage

\section{Introduction}%
\label{sec:intro}

Quantum Chromodynamics in the presence of an external magnetic field has recently attracted much attention because of its phenomenological relevance and of many interesting theoretical features, such as anomalous transport \cite{Kharzeev:2007jp, Fukushima:2008xe}, possibility of new phases in the QCD phase diagram \cite{Miransky:2015ava}, magnetic catalysis \cite{Gusynin:1994re, Gusynin:1994xp} and inverse magnetic catalysis \cite{Bali:2011qj,Bali:2011uf,Bali:2012zg,D'Elia:2012tr}. For comprehensive reviews for these and many other phenomena, see for example \cite{Kharzeev:2012ph, Kharzeev:2013jha, Miransky:2015ava}. 

Realization of these phenomena in nature typically requires very strong magnetic fields, $eB \sim \Lambda_\mathrm{QCD}^2$ and higher.  During non-central heavy ion collisions such large magnetic fields are believed to be generated by the spectator nucleons, and their magnitude  can reach up to $eB/\Lambda_\mathrm{QCD}^2 \sim 5-10$ at the time of the collision \cite{Skokov:2009qp, Tuchin1, Voronyuk:2011jd, Deng:2012pc, Tuchin2, McLerran:2013hla,Gursoy:2014aka}. Even though this magnetic field rapidly decays after the collision, it is still sufficiently strong at the time when the quark-gluon plasma forms, hence affecting the subsequent evolution of the plasma and finally the charged hadron production in these experiments \cite{Gursoy:2014aka}. Nuclear matter in strong magnetic fields also exists in other contexts such as the early universe \cite{Vachaspati:1991nm, Tashiro:2012mf,Semikoz:2007ti,Semikoz:2009ye,Semikoz:2012ka, Kahniashvili:2012uj,Dvornikov:2013bca} and neutron stars \cite{Duncan:1992hi}.  
 
In this paper we study the influence of external magnetic fields on the dynamics of the quark condensate in strongly interacting QCD-like theories. It is long known \cite{Gusynin:1994xp} that magnetic field has a constructive effect on the quark condensate at vanishing and low  temperatures. This is called the magnetic catalysis and the physical reason behind it can be understood as follows \cite{Gusynin:1994re, Gusynin:1994xp, Gusynin:1994va}: in the presence of a strong magnetic field, motion of the charged particles in directions transverse to $B$ are restricted due to the Landau quantization leading to an effective reduction from 3+1 to 1+1 dimensions. It is also well known that the IR dynamics in gauge theories are much stronger in lower dimensions, leading to a fortification of the quark condensate and catalysis of chiral symmetry breaking in the presence of magnetic fields. It came as a surprise therefore, when recent lattice studies found the opposite behavior at higher temperatures: for temperatures higher than a value slightly below the deconfinement crossover temperature, around $\sim 150$~MeV, the magnetic field is found to destroy the condensate~\cite{Bali:2011qj,Bali:2011uf,Bali:2012zg,D'Elia:2012tr}. This behavior, called the inverse magnetic catalysis, cannot be explained by perturbative QCD calculations, as these and various other effective models  predict catalysis instead \cite{Miransky:2015ava}. It is therefore believed to result from the strongly coupled dynamics in QCD around the deconfinement temperature. 

Among possible explanations of this phenomenon \cite{Miransky:2015ava}, the most promising one is the competition between the ``valence'' and the ``sea'' quark contributions to the path integral, which has been observed on the lattice~\cite{Bruckmann:2013oba, Bruckmann:2013ufa}. Here the valence quarks correspond to the quarks in the $\bar{q}q$ operator inside the path integral and the effect of $B$ through this contribution  always tend to catalyze the condensate, simply because $B$ increases the spectrum density of the zero energy modes of the Dirac operator. The ``sea'' contribution on the other hand comes from the quark determinant that describes fluctuations around the gluon path integral. $B$ and $T$ dependence of this contribution is more complicated and turns out to suppress the condensate around the deconfinement temperature. It is fair to say that a clear explanation of the puzzle of the inverse magnetic catalysis is still missing. 

We propose to study the problem in strong coupling and the limit of large number of colors ($N_c$) and flavors ($N_f$), using a realistic holographic bottom-up model for QCD based on \cite{ihqcd1, ihqcd2, ihqcd3, jk}. The model successfully incorporates breaking of the conformal symmetry and running of the gauge coupling, predicts realistic glueball and meson spectra \cite{ihqcd6, ikp1, ikp2, Iatrakis:2015rga} and fits very well the lattice results for temperature dependence of thermodynamic functions \cite{Panero:2009tv}. Since we work in the large $N_c$ limit and since the magnetic field couples the system only through the quarks, it is impossible to see phenomenon of inverse magnetic catalysis unless one also considers large number of flavors, i.e., the Veneziano limit \cite{Veneziano:1979ec}:   
\be
\label{vlim}
N_f, N_c \to \infty\,,\quad x={N_f \over N_c}=\text{fixed} \, , \quad \lambda=\frac{g_\mathrm{YM}^2 N_c}{8\pi^2} =\text{fixed} \, .
\ee
This is necessary, since the aforementioned ``sea'' quark contribution would be completely suppressed unless we also consider large $N_f$. The price one pays by taking the Veneziano limit is that the dual gravitational solution becomes much more complicated, since the backreaction of the flavor branes on the gravitational background has to be taken into account. A backreacted model was successfully constructed in \cite{jk} and the subsequent papers, \cite{aijk1, aijk2, alte, altemu, Iatrakis:2014txa, Alho:2015zua, Jarvinen:2015ofa, iksy, Arean:2016hcs}, and we shall use this model to study gravitational solutions with a finite magnetic field. It is important to stress that this holographic model successfully describes the dynamical chiral symmetry breaking at vanishing $B$. Holographic gauge theories in the presence of magnetic fields have been studied in several works in the past either at $N_f=0$ or  $N_f \ll N_c$, \cite{Zayakin:2008cy, Filev:2011mt, Erdmenger:2011bw, Preis:2012fh, Mamo:2015dea, Dudal:2015wfn, Rougemont:2015oea, Evans:2016jzo}, or with smeared backreacted flavor branes in the Veneziano limit~\cite{Jokela:2013qya}.

In the next section we present details of the model and discuss the numerical techniques we use to obtain the solutions. Section 3 contains the main results of our paper, in particular presence  of inverse magnetic catalysis in a particular range of  the parameter space. In section 4 we investigate the behavior of the condensate for varying number of flavors, by considering different values of $x$ in (\ref{vlim}). In the final section we discuss our findings in the light of the field theory and lattice QCD results discussed above. We leave the details of the equation of motion and the choice of potentials that define our theory in the appendices to simplify the presentation. 

\section{Holographic QCD in the Veneziano limit}
We consider an effective holographic model of QCD inspired by string theory and matched to low energy properties of QCD.  The dynamics can be separated in two sectors, the color and the flavor. The color part of the model is the Improved Holographic QCD (IHQCD) that describes the strong coupling dynamics of four dimensional Yang Mills in the large $N_c$ limit, \cite{ihqcd1, ihqcd2}. The low energy fields include the bulk metric and a real scalar, the dilaton, corresponding to the 't Hooft coupling.  The flavor part is constructed in a framework where flavors are introduced by $N_f$ coincident pairs of flavor branes and anti-branes \cite{Bigazzi:2005md,ckp}. The system is symmetric under the $U(N_f)_R\times U(N_f)_L$ flavor group. The lowest lying open string states on the flavor branes include a complex scalar field, the ``open string tachyon'', and the left and right gauge fields that  correspond to the left and right flavor currents respectively. The tachyon is dual to $ \bar q q $ operator and belongs to the bi-fundamental $(N_f, \bar N_f)$ representation of the flavor group. In the current work, we consider the full backreaction of the flavor sector to the glue in the Veneziano limit, equation (\ref{vlim}), 
and study the ground state of the system at finite temperature and magnetic field. The magnetic field is introduced in the flavor part of the model through the vector combination of the left and right gauge fields.

The glue action is,
\be
S_g= M^3 N_c^2 \int d^5x \ \sqrt{-g}\left(R-{4\over3}{
(\partial\lambda)^2\over\lambda^2}+V_g(\lambda)\right) \, .
\label{actg}
\ee
Here $\l=e^\phi$ is the exponential of the dilaton field. $M$ is the Planck mass in five dimensions. The Ansatz for the vacuum solution of the metric is
\be
ds^2= e^{2 A(r)}  \left( {dr^2 \over f(r)} -  f(r) dt^2+dx_1^2 +dx_2^2 +e^{2 W (r)}  dx_3^2\right) \,,
\label{bame}
\ee
where the anisotropy in $x_3$ direction is introduced because presence of the background magnetic field,  which we choose in the $x_3$ direction, breaks the rotational symmetry $SO(3)\to SO(2)$. The UV boundary lies at $r=0$ (where $A\to\infty$), and the bulk coordinate runs from zero to the horizon, $r_h$, where the black hole factor vanishes, $f(r_h)=0$. In the UV, $r$ is identified roughly as the inverse energy scale in the dual field theory. The dilaton potential, $V_g$, approaches a constant close to the boundary ($\l \to 0$). Its asymptotics in the IR ($\l\to \infty$) is $V_g\sim \l^{4\over 3}\sqrt{\log\l}$. This behavior is chosen to reproduce confinement, discrete glueball spectrum, linear Regge trajectories of glueballs and the thermodynamic properties of QCD \cite{ihqcd1, ihqcd2, ihqcd3, ihqcd6, ihqcd4, ihqcd5,  Iatrakis:2015sua}.

The flavor action was first proposed by Sen, \cite{sen}, in the study of a coincident brane-antibrane pair in flat spacetime. It was then employed in modeling the flavor sector of holographic QCD in \cite{ckp}, where it was shown to successfully reproduce the chiral symmetry breaking pattern and the low energy meson spectrum of QCD. It was generalized in the Veneziano limit by taking into account full backreaction of the flavor branes on the background geometry in \cite{jk}. These fully backreacted models are coined V-QCD. Sen's action at the vacuum with only real part of the tachyon nontrivial reads
\begin{align}
\label{actf}
S_f =-x\, M^3 N_c^2 \int d^5x V_f(\l,\t) \sqrt{- \mathrm{det}\left(g_{\m\n} + w(\l)\, V_{\m\n} + \kappa(\l)\, \partial_{\m} \t \,\partial_{\n} \t\right) } \,,\,\,
\end{align}
where $V_{\mu\nu}=\pa_{\mu}V_{\nu}- \pa_{\nu}V_{\mu}$ is the field strength of the bulk gauge field dual to $U(1)_{L+R}$ in the decomposition $U(N_f)_R\times U(N_f)_L \to SU(N_f)_R\times SU(N_f)_L \times U(1)_{L+R} \times U(1)_{L-R}$. We introduce the boundary magnetic field by choosing
\be
\label{vans}
 V_{\mu}= \left(0, -x_2 B/2, x_1 B/2,0,0 \right) \, ,
 \end{equation} 
 and set the other bulk gauge fields to zero. $\tau$ is the aforementioned tachyon field that we take in the diagonal form for simplicity. 
 The total action of the system is 
\be
S=S_g+S_f\,.
\label{action}
\ee
The tachyon potential has the form $V_f(\l,\t)=V_{f0}(\l) e^{- a(\l) \t^2} $ that has a minimum at $\tau \to \infty$. Presence of this field is crucial to the model since it serves as an order parameter of the chiral symmetry breaking  \cite{ckp}.  As the tachyon potential is minimized  by requiring $\tau$ diverge in the IR of the geometry, then, the brane - anti-brane pairs condense and, if the geometry is confining, chiral symmetry breaks at zero temperature \cite{ckp}. Above a certain temperature however the profile with minimum energy becomes $\tau=0$ restoring chiral symmetry. 

The functions $V_{f0}(\l)$, $a(\l)$, $\k(\l)$ and $w(\l)$ need to satisfy several constraints. First we need to fix the behavior in the UV, i.e., at weak coupling $\l \to 0$. In this regime the holographic model is not expected to be reliable, however we can still make a choice that guarantees the best possible UV ``boundary conditions'' for the more interesting IR physics, and choose the potentials consistently with QCD perturbation theory~\cite{ihqcd1, ihqcd2, jk}. In particular,  we determine $V_{f0}$ and $\k/a$ near the UV boundary such that the holographic RG flow of the dilaton and the tachyon matches the perturbative RG flow of the coupling and the quark mass in QCD. 
Consequently, the leading boundary behavior of the bulk scalar fields is given by
\begin{align}
\label{lauv}
\l(r) &\simeq - \frac{b_0}{\log \Lambda r} & \\
\tau(r) &\simeq m_q r(-\log \Lambda r)^{- \rho}+\langle {\bar q} q \rangle r^3 (-\log \Lambda r)^{\rho}&
\label{tauuv}
\end{align}
where $b_0$ is the leading coefficient of the QCD $\beta$-function and the power $\rho$ is also matched to the coefficients of the anomalous dimension of ${\bar q} q$ and the QCD $\beta$-function (see~\cite{jk,aijk2} for details). In this work we only consider massless quarks so the non normalizable mode of the tachyon solution vanishes. 

The UV energy scale $\Lambda$ in~\eqref{lauv} and~\eqref{tauuv} is identified with $\Lambda_\mathrm{QCD}$ on the field theory side up to a proportionality constant. We stress, however, that while this proportionality constant is typically $\mathcal{O}(1)$ it does not need to be very close to one. As is the case for lattice QCD, the matching of the energy scales of the holographic model and real QCD needs to be done by comparing the values of some physical parameter, such as the pion decay constant, the mass of the $\rho$-meson, or the critical temperature of the confinement-deconfinement transition. For the potentials which we shall use here, this matching typically leads to the value of $\Lambda$ being around 1~GeV.

The functions $V_{f0}(\l)$, $a(\l)$, and $\k(\l)$ are constrained also in the IR ($\l \to \infty$) by requiring that they reproduce the expected dynamics in the flavor sector, such as the phase diagram of the theory with varying $x$, $T$ and chemical potential and the properties of meson spectra, \cite{aijk1, aijk2, alte, altemu, ikp1, ikp2, Alho:2015zua, Jarvinen:2015ofa}.  In the present work, we use the choice for these potentials constructed in \cite{altemu}. We present this choice of potentials in Appendix~\ref{app:potentials}. 

The most important coupling function in our study is $w(\l)$ because it couples the electromagnetic sector of the theory to the gluon dynamics, hence its shape strongly affect electromagnetic properties of dual theory that we are interested. Its asymptotic dependence on $\l$ is not strongly constrained by studies referred above. The most natural expectation is that $w(\l)$ and $\kappa(\l)$, i.e., both couplings in the square root factor of the DBI action, have similar asymptotics both in the UV and in the IR. This assumption is consistent with the UV behavior of the two point function of the flavor vector current and asymptotics of the meson spectra~\cite{ikp2, aijk2}. The potential $w$ also plays an important role in determining the transport properties of the Quark-Gluon Plasma, such as its conductivity and the diffusion constant \cite{Iatrakis:2014txa}. Hence, it is also a major factor in the  calculation of the spectrum of emitted photons in the QGP phase of heavy ion collisions \cite{iksy}. Assuming that $\kappa$ and $w$ have similar IR asymptotics lead to reasonable physics and yield good fits to the experimental data also in these studies.

Motivated by these earlier studies, we therefore make a choice for $w$ for which it has the same asymptotics as $\kappa$: 
\be \label{wltext}
 w(\l) = \kappa(c\, \l) \ ,
\ee
where $c$ is a parameter. We will see that judicious choices of $c$ lead to interesting phenomena such as the inverse magnetic catalysis.

\section{Numerical Results}\label{sec:numerics}

Solving Einstein's equations derived from the action (\ref{action}), we extract the phase diagram of the model as a function of the temperature and the magnetic field. Furthermore, we determine the chiral condensate and show that for particular choices of $w(\lambda)$ the model exhibits inverse magnetic catalysis in qualitative agreement with the lattice results, \cite{Bali:2011qj,Bali:2011uf,Bali:2012zg,D'Elia:2012tr}. In more detail, we solve Eqs. (\ref{backeq}, \ref{laeq}, \ref{tacheq}) of Appendix~\ref{app:eoms} by shooting from the horizon towards the boundary. For each bulk solution we fix the non-normalizable solution of the fields close to the boundary and read the normalizable asymptotics. Then we determine the vacuum expectation values of the dual field theory operators, i.e., the temperature, magnetic field and chiral condensate of the dual field theory state. 

As shown in \cite{alte, altemu}, the holographic V-QCD model often has a first order confinement-deconfinement transition at $T_d$, and a separate second order chiral transition at $T_{\chi}$, where $T_{\chi}>T_d$. For $T<T_d$ the thermodynamically dominant solution is the thermal gas geometry corresponding to a confined and chirally broken field theory state. For $T_d<T<T_{\chi}$, the geometry is a black hole with a tachyon hair, so that the dual state is deconfined and chirally broken. Finally, for $T>T_{\chi}$, the geometry is a tachyonless black hole, and therefore the dominant phase is deconfined and chirally symmetric. 
Such separate transitions were seen at $B=0$ for values of $x$ close to the conformal transition of QCD for all studied potentials, but for the particular choice of potentials of Eqs. (\ref{Vf0SB}) and~\eqref{kappaa} this behavior is seen even at low values of $x$, down to $x \simeq 1$.

\begin{figure}[!tb]
\begin{center}
\includegraphics[width=0.49\textwidth]{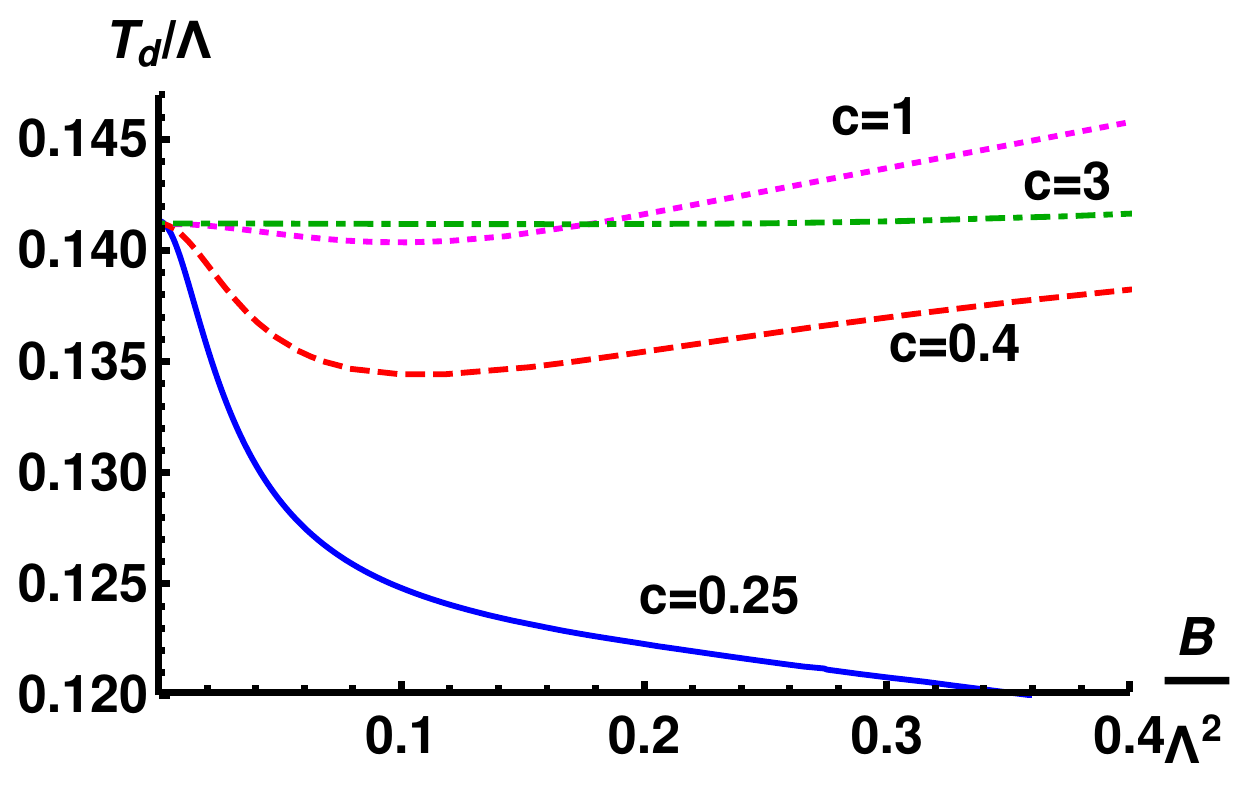}
\includegraphics[width=0.49\textwidth]{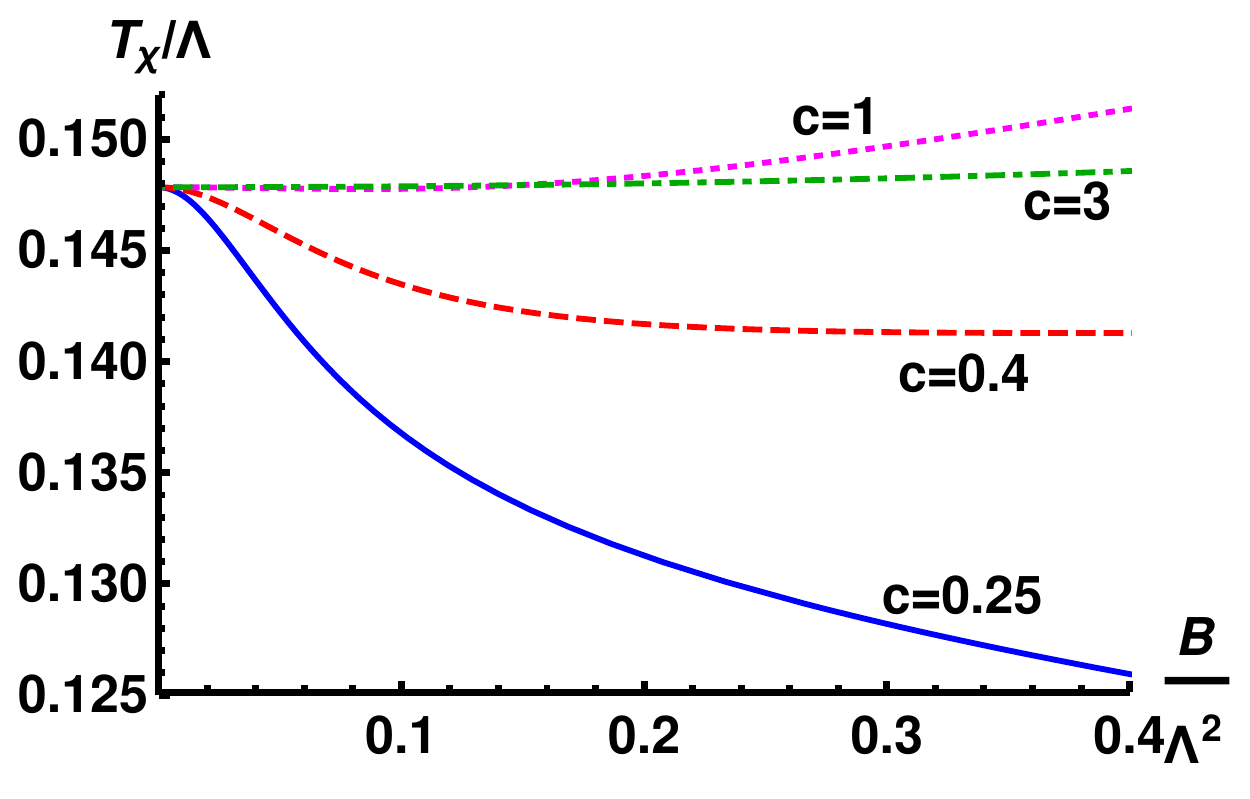}

\end{center}
\caption{The confinement-deconfinement transition temperature $T_d$ (left) and the chiral transition temperature $T_\chi$ (right) as a function of the magnetic field for different values the parameter $c$, for zero quark mass, $m_q=0$ and for $x=1$. $T_d$ and $B$ are measured in units of the energy scale $\Lambda$.}
\label{fig:Tdc}
\end{figure}

As we shall see, adding a finite background magnetic field does not drastically change this phase structure. The coupling function $w(\lambda)$ plays an important role on the dependence of the transition temperatures $T_d$ and $T_\chi$ on $B$, since it controls the interaction of the medium with the magnetic field. Hence, we study the transition temperatures and the condensate as function of $B$ for different choices of $w(\lambda)$ parametrized by the parameter $c$ appearing in Eq.~(\ref{wltext}). We start by studying the $B$ dependence at moderately low values of $B$, but still high enough for the backreaction to be important. The left plot in Fig. \ref{fig:Tdc} shows the deconfinement temperature as a function of $B$ for different $c$. We observe that for sufficiently small values of $c$, the transition temperature $T_d$ decreases as a function of $B$. For larger values, i.e., for $c \gtrsim 1$, the dip in the deconfinement temperature is suppressed and growing behavior with $B$ dominates\footnote{We find numerically that the dip does not disappear, but becomes extremely weak as $c$ increases. $T_d$ grows with $B$ also for $c=3$ even though this is barely visible in figure \ref{fig:Tdc}. The growth becomes more pronounced at higher $B$ for this value of $c$.}. All dimensionful quantities are measured in units of $\Lambda$, that is an energy scale of the model which appears as an integration constant in~\eqref{lauv} and~\eqref{tauuv}. 

\begin{figure}[!tb]
\begin{center}
\includegraphics[width=0.59\textwidth]{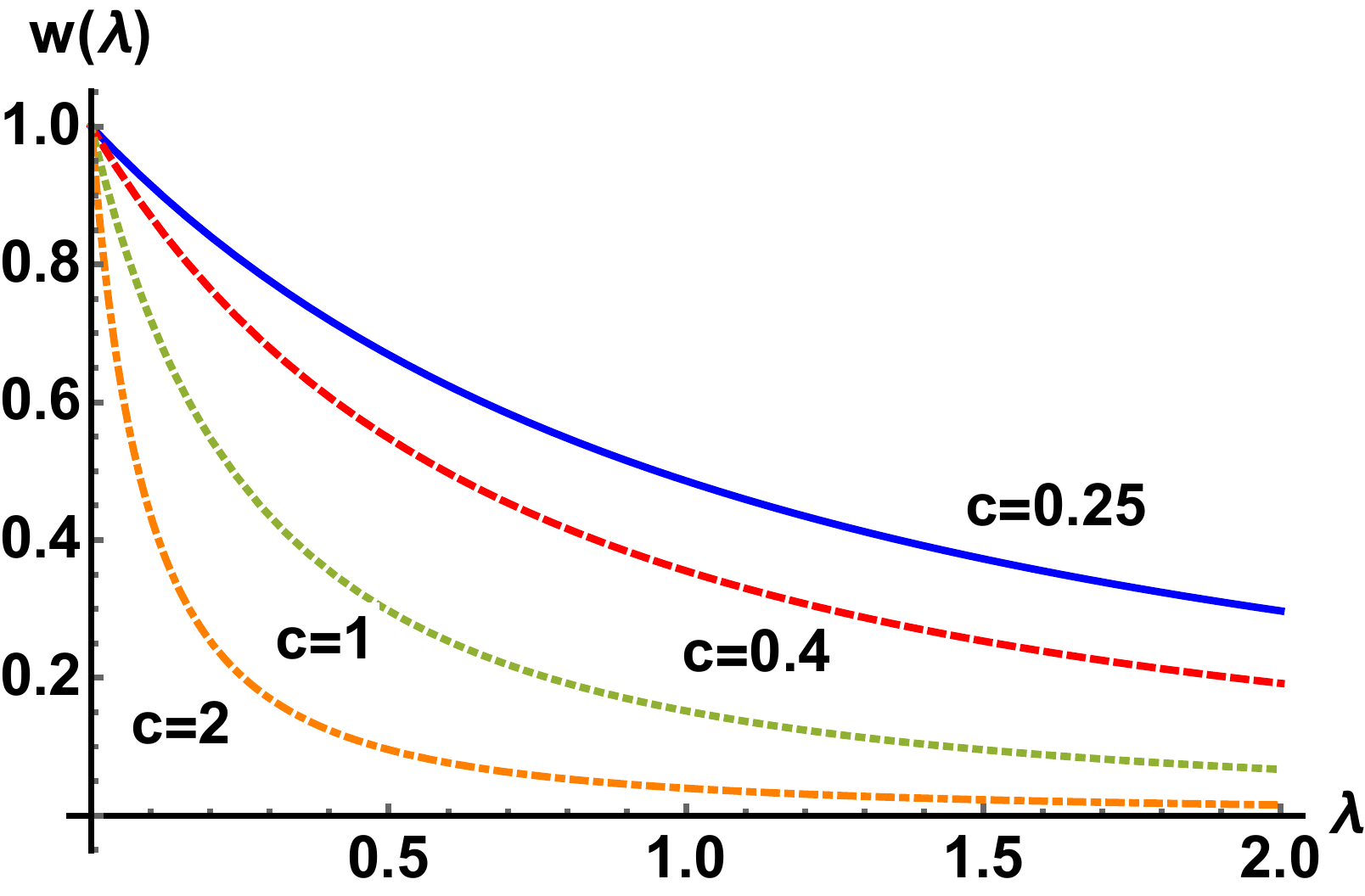}
\end{center}
\caption{Dependence of the function $w(\lambda)$ on the parameter $c$ for $x=1$. The curves are for $c=0.25$, $0.4$, $1$ and $3$. We note that increasing $c$ suppresses $w(\lambda)$.}
\label{figw}
\end{figure}

The right plot in Fig. \ref{fig:Tdc} depicts the chiral transition temperature as a function of $B$ for different values of $c$. We note that for small values of $c$, i.e., $c<0.4$, the chiral transition temperature is a decreasing function of $B$,  a fact that signals inverse magnetic catalysis. The function $w(\l)$ takes larger values for smaller $c$ as shown in figure \ref{figw}. This means that the coupling of the magnetic field to the glue dynamics, i.e., the dilaton, becomes stronger for smaller values of $c$. As a result, we qualitatively expect that the effect of the quarks to the transition temperature becomes more important  and eventually leads to inverse magnetic catalysis. This argument is in qualitative agreement with findings in \cite{iksy}, where it is shown that a large $w(\l)$, compared to the $c=1$ case matches better the lattice result for the electric conductivity of QGP at vanishing $B$. A detailed phenomenological matching of the model to low energy QCD is a subject we leave for future work, but it is reassuring that the results of our preliminary analysis here are in qualitative agreement with electromagnetic properties of QGP. The choice $c=0.4$ seems to correctly reproduce the qualitative features observed in the lattice studies. 

\begin{figure}[!tb]
\begin{center}
\includegraphics[width=0.59\textwidth]{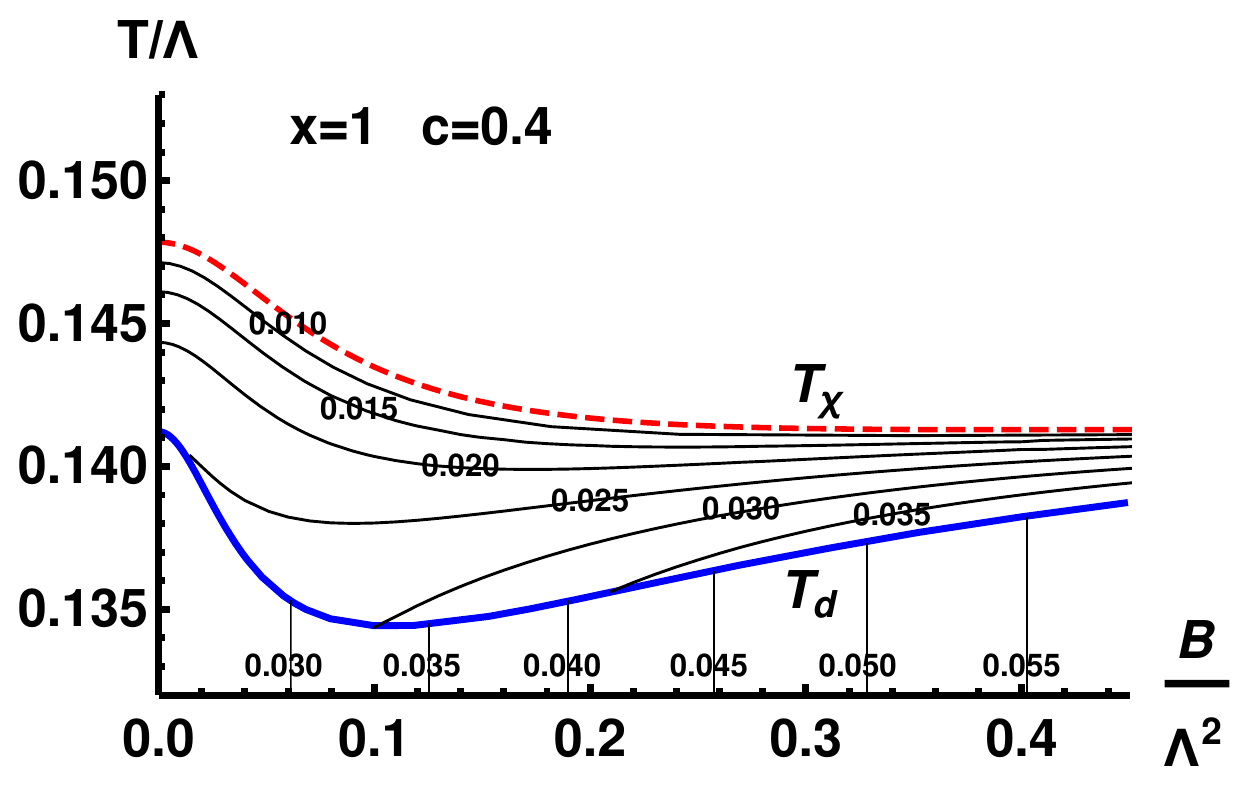}
\end{center}
\caption{Curves of constant $\langle \bar q q \rangle $ on the  $T-B$ plane, in units of $\Lambda$, for $c=0.4$, zero quark mass, and $x=1$. The labels on the curves correspond to the value of $\langle \bar q q \rangle / \Lambda^3$. Below the blue curve, corresponding to the deconfinement transition, the condensate is independent of temperature, hence the lines are straight. Moreover, as the chiral transition (the red dashed line) is approached from below the value of the condensate approaches zero.}
\label{fig:conqq}
\end{figure}

Since we have included the dynamics of the flavor sector in the full backreacting regime of $N_f\sim N_c$, we are able to explicitly compute the quark condensate using our model. Using the standard holographic techniques, we set the non-normalizable boundary solution of the tachyon to zero, which corresponds to zero quark mass, and then read numerically the value of the condensate form the normalizable solution of Eq. (\ref{tauuv}). In Fig. \ref{fig:conqq}, curves of constant chiral condensate are plotted. Higher curves correspond to lower values of the condensate, and finally the red dashed line is the chiral transition, along which the condensate vanishes. Hence, we observe that the condensate is a decreasing function of $B$ at fixed temperature, that is indeed the phenomenon of inverse magnetic catalysis. The reason for the straight contours for the condensate below the blue curve is that this phase correspond to the thermal gas background in the holographic dual, for which the temperature dependence of all thermodynamic functions is suppressed as $1/N_c$ in the large-N limit.

\begin{figure}[!tb]
\begin{center}
\includegraphics[width=0.49\textwidth]{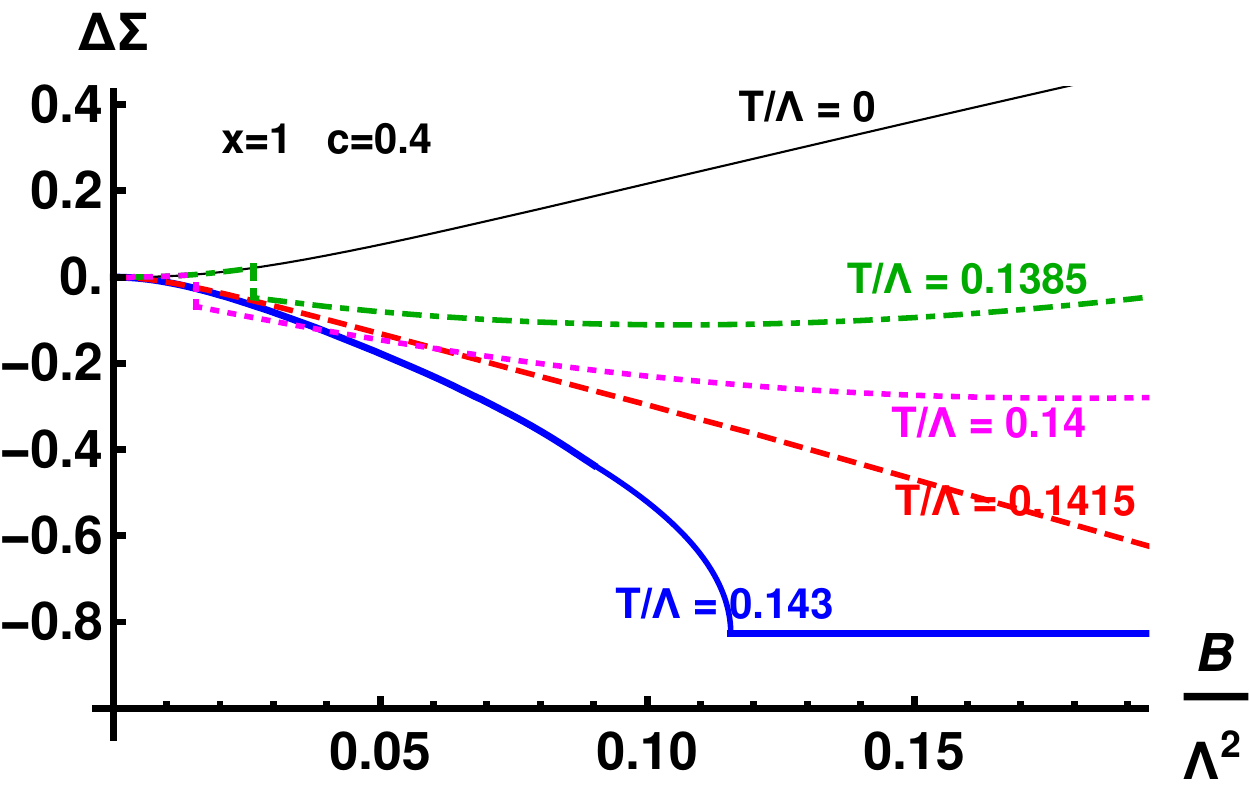}
\includegraphics[width=0.49\textwidth]{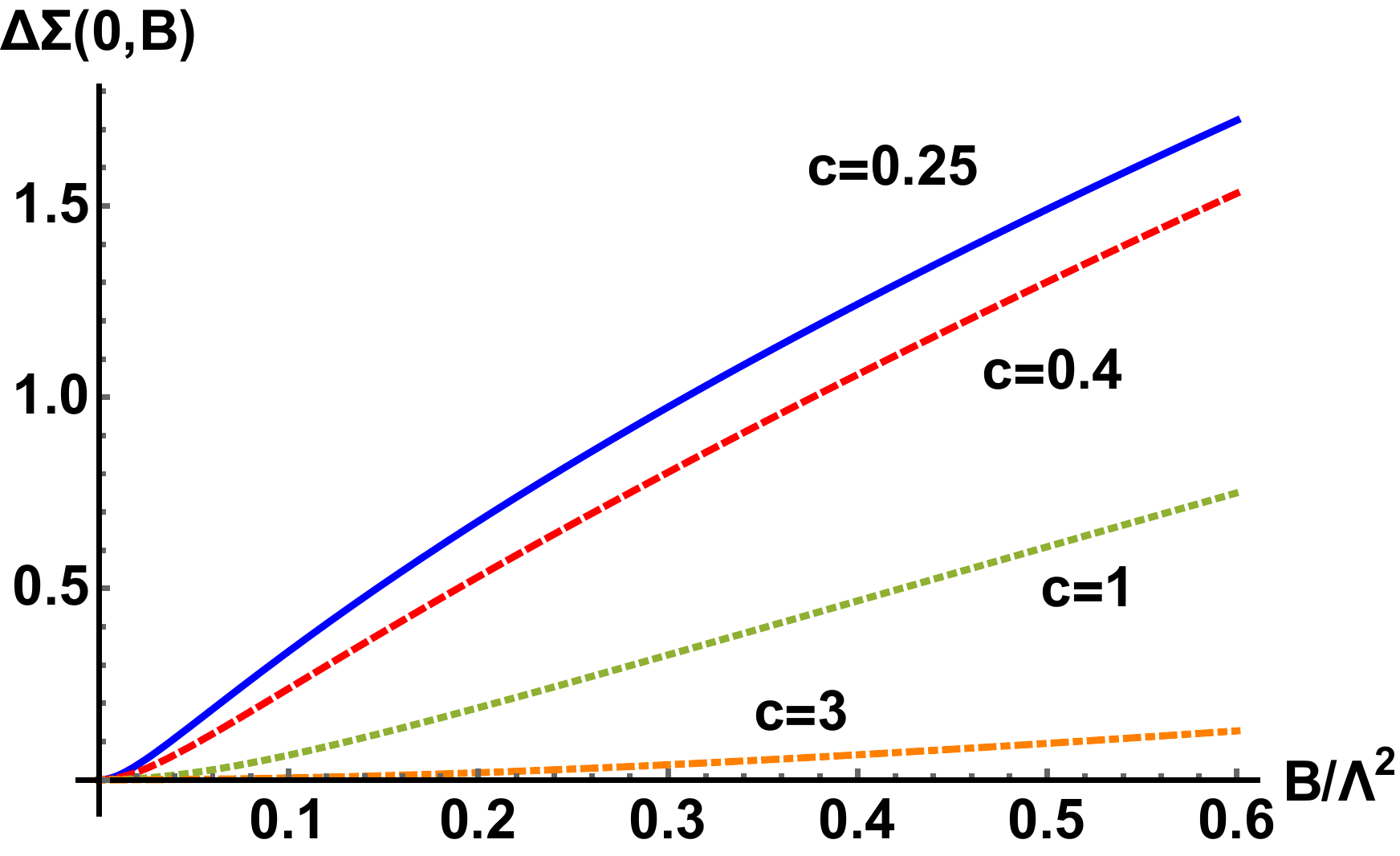}
\end{center}
\caption{{\it Left:} The normalized variation of the chiral condensate, $\Delta \Sigma(B,T)$, as a function of $B$ for constant $T$, $c=0.4$, zero quark mass, and $x=1$.  {\it Right:} The normalized chiral condensate in the confined phase ($T=0$) as a function of the magnetic field at zero quark mass and $x=1$ and for different values of the parameter $c$. }
\label{fig:Dqq}
\end{figure}

The chiral condensate is not invariant under the renormalization group flow. A renormalization group invariant combination reads
\be
\Sigma(T,B)= \frac{\langle \bar q q \rangle (T,B)}{\langle \bar q q \rangle (0,0)}= {1 \over \langle \bar q q \rangle (0,0) } \left(  \langle \bar q q \rangle (T,B)-\langle \bar q q \rangle (0,0) \right)+1 \, ,
\ee
that is dimensionless. The change due to the magnetic field is then defined as
\be
\Delta \Sigma(T,B) =\Sigma(T,B)-\Sigma(T,0) \,.
\ee
This difference $\Delta \Sigma $ is plotted for V-QCD in the left plot of Fig. \ref{fig:Dqq} for vanishing quark mass and $c=0.4$. We find very good qualitatively agreement with the lattice results 
\cite{Bali:2012zg}. One main difference, is the fact that in the Veneziano limit, QCD has a first order confinement-deconfinement transition, hence the condensate jumps at this point. However, the picture is very similar to the $N_c=N_f=3$ case. At zero temperature there is magnetic catalysis. For larger temperatures $\Delta \Sigma$ increases for small $B$ and then
it jumps and starts decreasing for higher $B$. For even larger temperatures, $\Delta \Sigma$ is a monotonically decreasing function of $B$, at least in the range of $B$ which is plotted. We also observe that for intermediate temperatures ($T/\Lambda=0.1385,0. 14$), $\Delta \Sigma$ starts to increase for larger values of $B$. For $T/\Lambda=0.143$, the condensate hits the chiral transition at $B/\Lambda^2=0.116$, so for larger values of $B$ it is zero.

The right plot in Fig. \ref{fig:Dqq} depicts the normalized condensate as a function of $B/\Lambda^2$ in the confined phase of the model for the choices $x=1$ and $m_q=0$ and for various values of the parameter $c$. We note that for $B/\Lambda^2 \ll 1$ the normalized condensate behaves as $\Delta \Sigma(0,B)=D_{\bar qq}(c) B^2 + \cdots$, where the parameter $D_{\bar qq}$ depends on $c$. We numerically determine $D_{\bar qq}(0.25)\simeq 124\, , \,\, D_{\bar qq}(0.4)\simeq 58\, , \,\,D_{\bar qq}(1)\simeq 5.3\, , \,\, D_{\bar qq}(3)\simeq 1.8$. We further note that the slope  $D_{\bar qq}$ decreases with increasing $c$. This is very much expected, since the coupling of the magnetic field to the plasma is controlled by the function $w$ that is more pronounced for smaller values of $c$ as shown in figure \ref{figw}.

Finally, we find linear dependence on $B$ for large $B$ as a consequence of how $B$ enters the equations of motion in Appendix~\ref{app:eoms}: it enters through the combination
\be \label{Qdeftext}
 Q(r)=\sqrt{1+w(\l)^2 B^2 e^{-4 A(r)}} \,,
\ee 
that is indeed linear $Q(r)  \simeq  e^{-2 A(r)} w(\l) B$ for large magnetic fields. 

Some of the important thermodynamic observables of the QGP medium related to the magnetic field are the magnetization $M_B$ and the magnetic susceptibility $\chi_B=\left. {M_B \over B} \right|_{B=0} $. They have  been computed in the lattice for 2+1 flavors at $B=0$ in \cite{Bonati:2013vba}. The susceptibility in our holographic model is given by 
\be 
 \chi_{B}=-{1\over V_4} \left.  { \partial^2 {\mathcal S_E^\mathrm{on-shell}} \over \partial B^2} \right |_{B=0}\,, 
\ee
where ${\mathcal S_E^\mathrm{on-shell}}$ is the Euclidean on shell action of the model. Inserting here the expression for the flavor action~\eqref{actf} we obtain
\be
\chi_B=\, M^3 N_c^2 \, \int_{r_h}^{r_\epsilon} dr\,  x\, V_f(\lambda,\tau) w(\l,\t)^2 e^{A(r)+W(r)} G(r) \,,
\ee
where
\be
 G(r) = \sqrt{1 + e^{-2 A(r)}\kappa(\l,\tau) f(r) (\partial_r \t(r))^2} \, , 
\ee
$r_h$ is the location of the horizon and $r_\epsilon$ is a cut-off near the boundary. The magnetization of the ground state is (at any value of $B$) 
\begin{align}
 M_{B}&=-{1\over V_4} { \partial {\mathcal S_E^\mathrm{on-shell}} \over \partial B} \nn \\
 &=\, M^3 N_c^2 \, \int_{r_h}^{r_\epsilon} dr\,  B\, x\,  V_f(\lambda,\tau) w(\l,\t)^2 e^{A(r)+W(r)} {G(r)\over Q(r)}\,.
\end{align}
Both the susceptibility and the magnetization diverge at the boundary and have to be renormalized appropriately. We do this here by subtracting their values for reference (thermal gas) solutions at $T=0$.

\begin{figure}[htp]
\centering
\includegraphics[width=0.59\columnwidth]{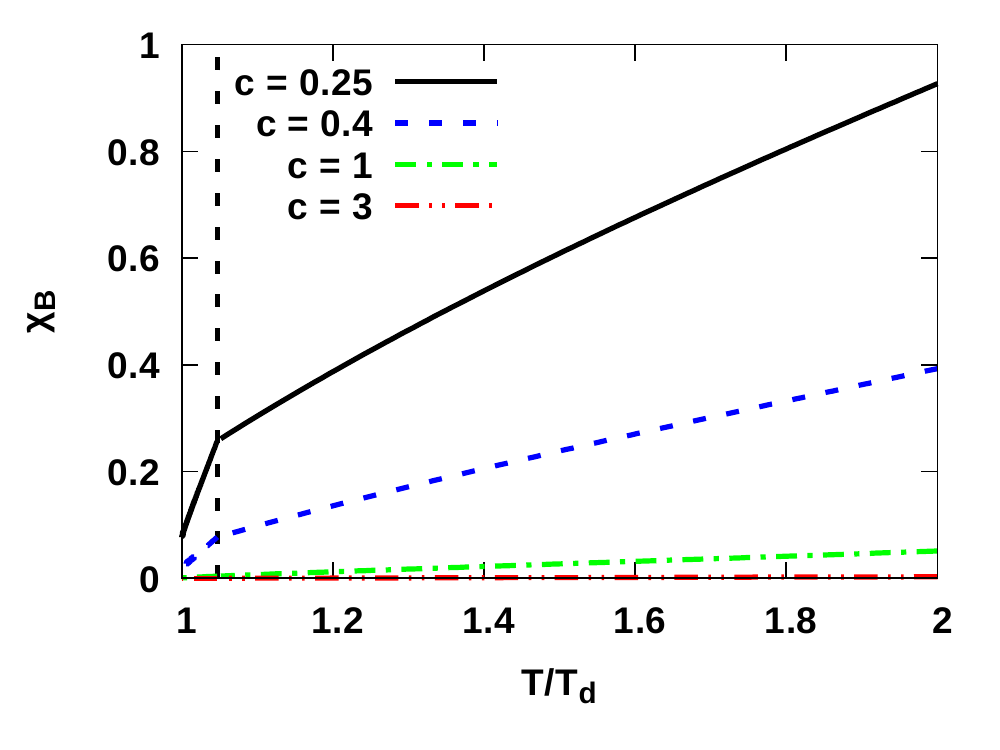}
\caption{\label{suscx1}Magnetic susceptibility as a function of temperature for various values of $c$ for $x = 1$. The chiral transition is indicated by the vertical dashed line.}
\end{figure}

In figure \ref{suscx1}, we show the magnetic susceptibility $\chi_B$ as a function of temperature for different values of $c$.
The plot shows that smaller $c$ leads to larger values of $\chi_B$.
This is expected by the following argument. Since the function $w$ that controls the coupling of the magnetic field to the plasma is more pronounced for smaller $c$, we expect the effect of quarks become more important, yielding a stronger inverse magnetic catalysis, in other words, a steeper decrease in $T_d$ around $B=0$. Now, because at the deconfinement transition near $B = 0$ we have $\mathrm{d}F = 0$ and hence $\mathrm{d}T_d/\mathrm{d}B = -\chi_BB/S$, a stronger decrease in $T_d$ with $B$ results in a larger positive value of $\chi_B$.
Another observation is appearance of kinks at the chiral transition $T = T_\chi$ that is different from the deconfinement transition ($T_\chi>T_d$) for $x=1$.

\section{Varying number of flavors}

\begin{figure}[htp]
\centering
\includegraphics[width=\columnwidth]{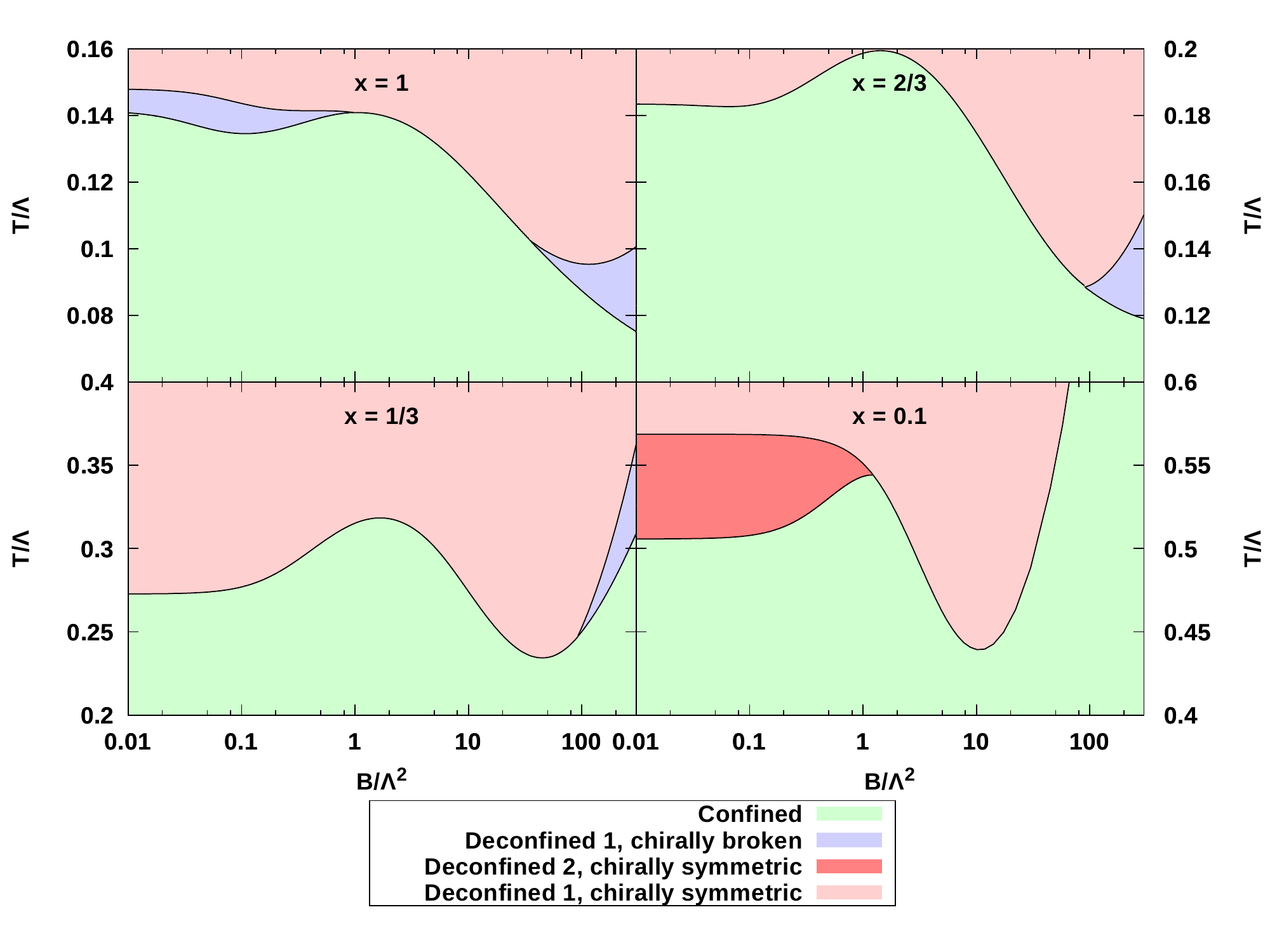}
\caption{\label{fig:varyingx}The phase diagram of our model for various values of $x$ and for $c = 0.4$. $T$ and $B$ are measured in units of $\Lambda$.}
\end{figure}

In our holographic model both the number of colors $N_c$ and the number of flavors $N_f$ are taken to be infinite with their ratio $x = N_f/N_c$ fixed.
By varying $x$ then it should be possible to study the influence of the quark sector on (inverse) magnetic catalysis. It is also interesting to investigate whether the phase diagram show additional features in the regime with $B/\Lambda^2 \gg 1$ for different values of the ratio $x$.
We address these questions  in figure \ref{fig:varyingx}, where the phase diagrams in the $(T,B)$-plane are shown for different values of $x$. In this plot we also extend the range of $B$ to much larger values than in Section~\ref{sec:numerics}.

Several interesting features arise in these diagrams. First, we observe inverse magnetic catalysis kicking in around $B/\Lambda^2 \sim 1$ on for all the choices of $x$. Moreover, we observe magnetic catalysis taking over for larger values of $B$ for smaller choices of $x = 0.1$ and $1/3$.
For $x = 2/3$ and $1$ there are also hints that the deconfinement transition might start increasing again at large $B$, but numerics in that region is not stable enough to assert this with certainty.
Another interesting feature is the reappearance of the deconfined chirally broken phase at large $B$. 
This can be seen in all diagrams for $B/\Lambda^2 \gtrsim 100$, except\footnote{In this case, there are hints of a triple point around $B/\Lambda^2 \sim 100$, but the numerics is not stable enough to confirm this with certainty.} for $x = 0.1$.
Finally we obtain an extra deconfined, chirally symmetric phase  for $x = 0.1$, that is separate from the other deconfined chirally symmetric phase by a first order transition.
This additional phase transition has also been observed at $B = 0$ in \cite{alte}, and is discussed in more detail there. 

\begin{figure}[htp]
\centering
\includegraphics[width=0.59\columnwidth]{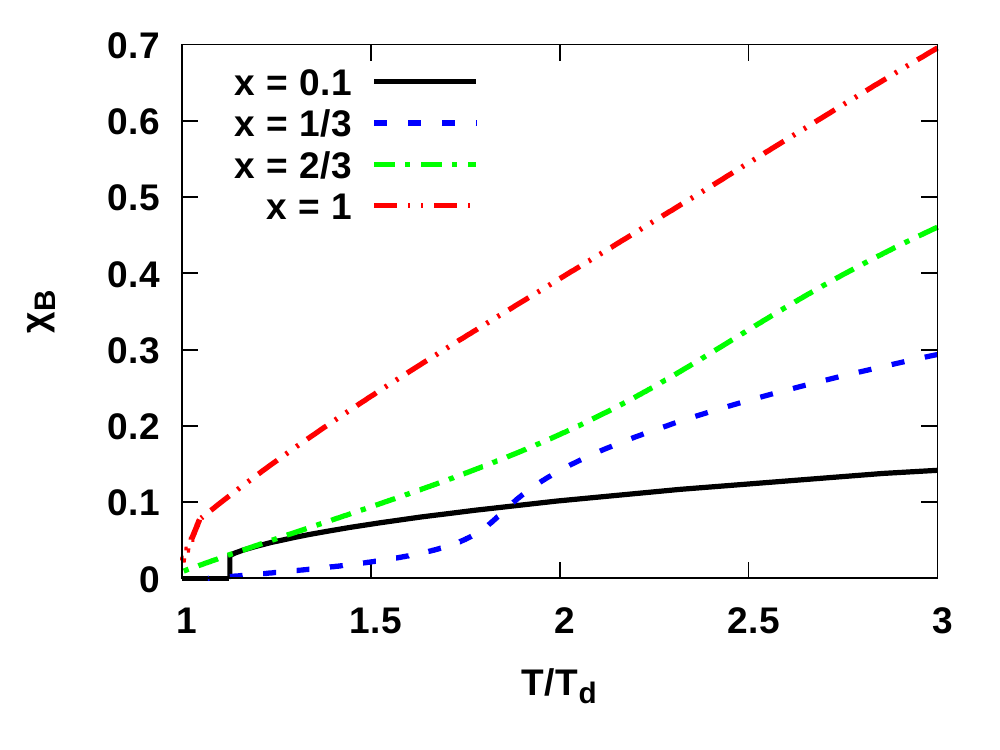}
\caption{\label{suscc04}Magnetic susceptibility as a function of temperature for different values of $x$ for $c = 0.4$.}
\end{figure}

It is also interesting to study dependence of the magnetic susceptibility on $x$. We plot this dependence in Fig. \ref{suscc04}.
One can see here that the shape of the curves varies non-trivially with $x$. First of all we observe that the susceptibility becomes larger for larger values of $x$. This is obvious as the susceptibility controls the reaction of the plasma, in particular its magnetization to $B$, and this reaction is stronger for larger number of flavors.  More interestingly, we observe that,  an inflection point forms as $x$ is decreased, as can clearly be seen in the curve for $x=1/3$. This inflection point moves toward smaller values of $T$ with decreasing $x$, which eventually turns into a jump in susceptibility at around $x = 0.1$. The jump can be traced back to the phase transition between the two chirally symmetric deconfined phases (red and purple) shown in Fig. \ref{fig:varyingx} for $x=0.1$.

Finally, we would like to identify the physical reason behind inverse magnetic catalysis using our holographic model. As we argue below studying the behavior of the condensate on $x$ is helpful in this quest. As mentioned in the Introduction, magnetic field influences the chiral condensate through two separate sources as shown in the lattice studies \cite{Bruckmann:2013oba, Bruckmann:2013ufa}. The first source is the explicit coupling of the quarks in the operator $\langle \bar q q \rangle $ to the magnetic field.  This direct coupling is termed as ``valence'' quarks in  \cite{Bruckmann:2013oba, Bruckmann:2013ufa}, and it is argued to strengthen the condensate, hence causing magnetic catalysis. The second source is the indirect dependence of the condensate expectation value on the magnetic field that arises from the the quark determinant in the path integral. This source, termed as the ``sea'' quarks, arise from the  the gluon path integral, therefore directly related to the glue physics. It was argued in \cite{Bruckmann:2013oba, Bruckmann:2013ufa} that this backreaction of the glue physics is to cause the inverse effect and it dominates over the catalysis created by valence quarks, in the large B regime. 

We can also identify two separate sources of B to the condensate in our holographic model. The first is the explicit dependence of the tachyon field equation  Eq.~(\ref{tacheq}) on B. This explicit dependence is present in the function $Q(r)=\sqrt{1+w(\l)^2 B^2 e^{-4 A(r)}}$ both in the linear $\sim \t'$ and in the nonlinear $\sim \t'^3$ term in this equation. The second source is the indirect effect coming from the modification of the background functions, the metric $g_{\mu\nu}$ and the dilaton $\lambda$ by presence of B. The change in these background functions due to B also affects the tachyon equation in an indirect manner. It is very tempting to identify the former explicit dependence with the valence, and the latter, implicit dependence with the sea quarks.  

In order to test this idea one can try to isolate one of these effects. First, let us consider the large B limit of the tachyon equation in the confined phase. We note that in the large B limit $Q(r) \sim  e^{-2 A(r)} w(\l) B $ and the tachyon equation Eq.~(\ref{tacheq}) simplifies as
\begin{align} \label{tacheqBB}
&\t''-{e^{2 \, A} \, G^2 \over f\, \kappa(\l)}{\partial_{\t} \log \, V_f (\l,\t)} + \left(A'+W' +{f' \over f} + \l' \partial_{\l} \log(V_f(\l,\t) \, \k(\l)) +\l'  \partial_{\l} \, \log \, w(\l) \right) \,\t'  \nn \\
&+\, e^{-2 \, A} \, f \, \k(\l) \, \left( W' +{1 \over 2} \,{ f' \over f} + 2 A'  +{1 \over 2}\, \l' \, \partial_{\l} \log \,( \kappa(\l)  \,V_f(\l,\t)^2 ) + \l'  \partial_{\l} \, \log \, w(\l) \right) \t'^3 =0 \, . \nn
\end{align}
The explicit dependence on the magnetic field $B$ cancels out in this limit. This is very much in accord with our assertion because according to the lattice study in the deconfined phase and in the large B region, it should be the glue physics, the sea quarks controlling the behavior of the condensate. Indeed, in this limit the only dependence of the tachyon equation (\ref{tacheq}) on B comes from the implicit dependence in the metric functions that we want to match with the sea quark effect. 

\begin{figure}[htp]
\centering
\includegraphics[width=.49\textwidth]{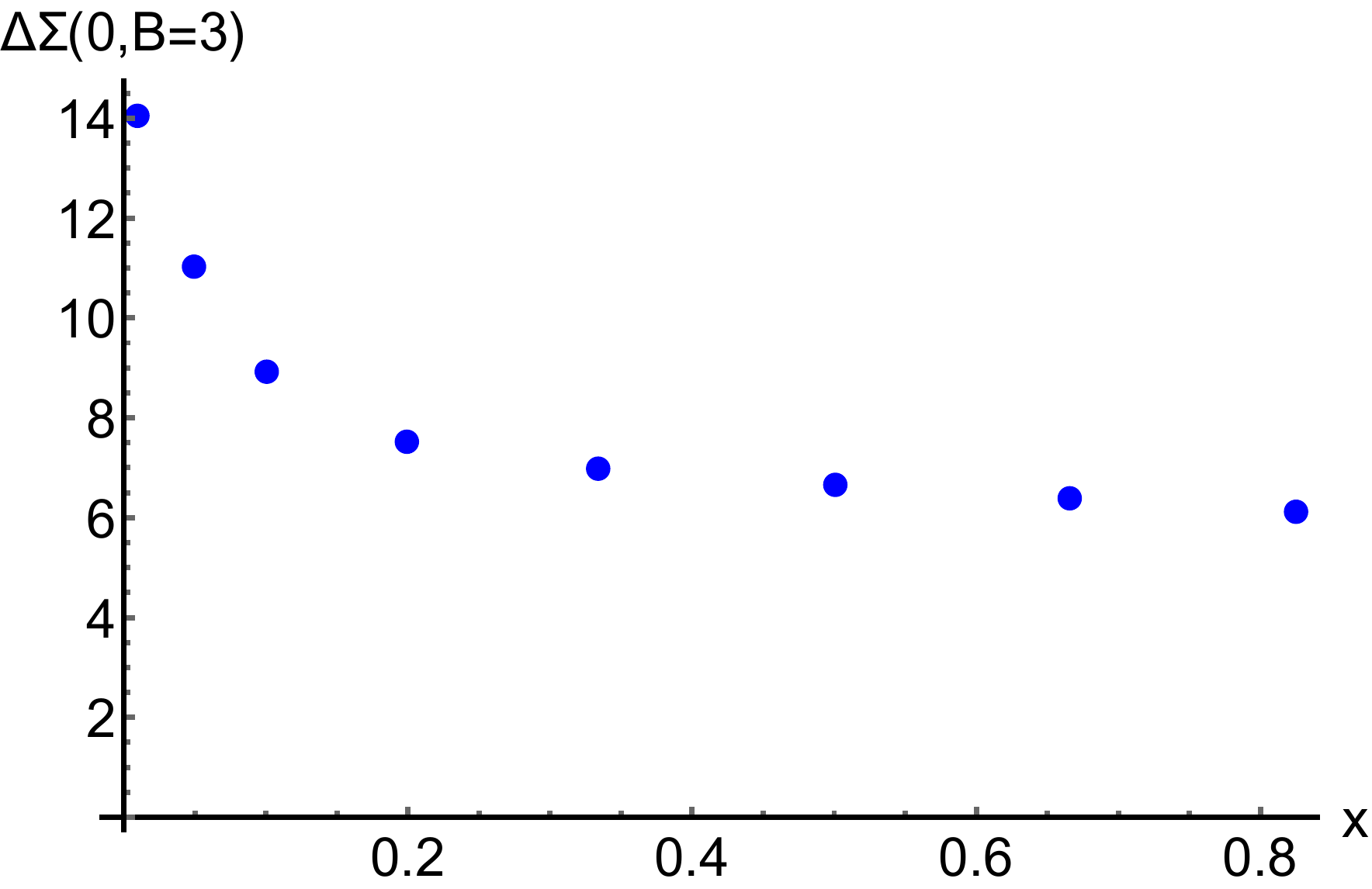}
\includegraphics[width=.49\textwidth]{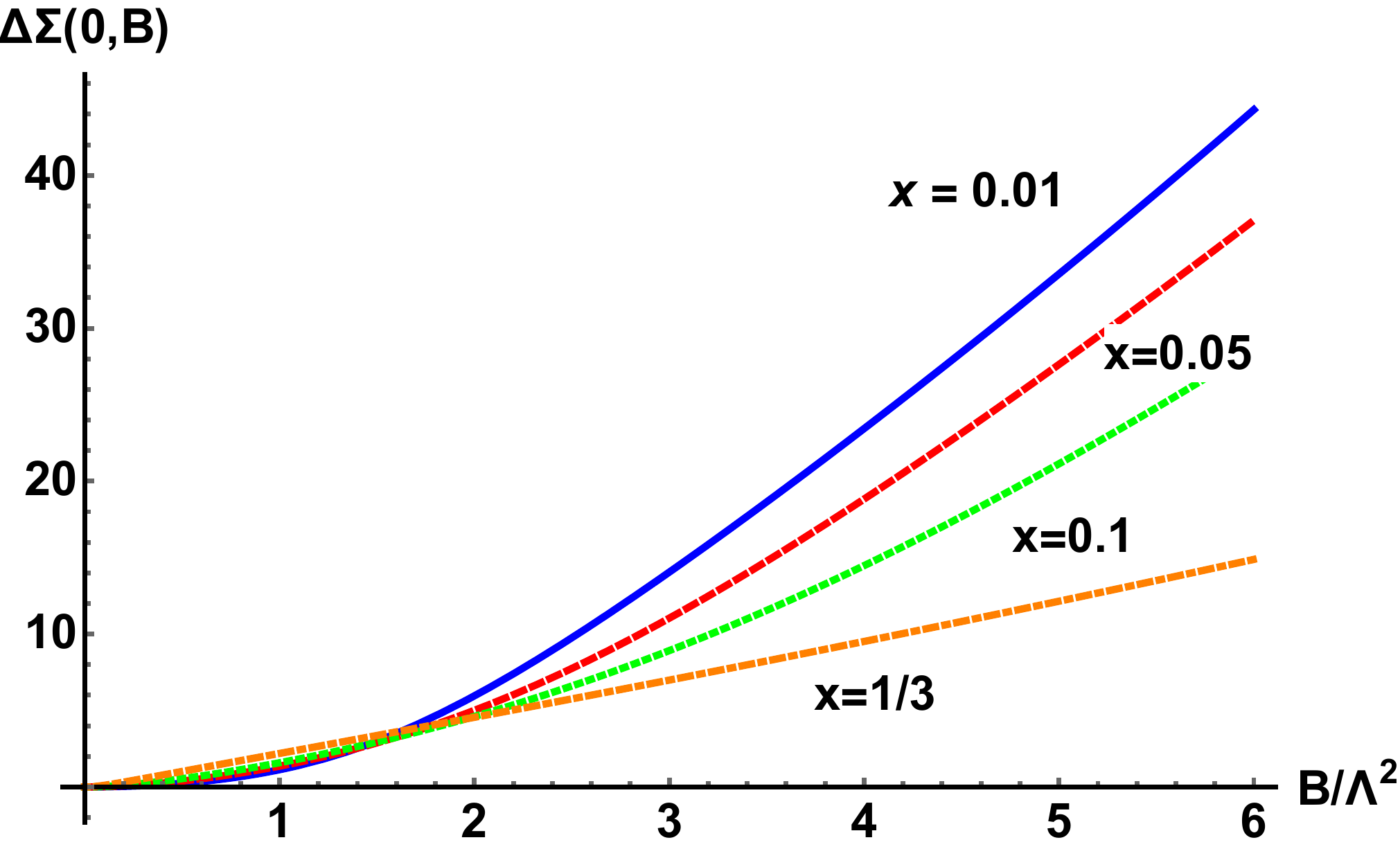}
\caption{{\it Left:} The normalized condensate in the confined phase ($T=0$) for $B/\Lambda^2=3$ as a function of $x$. Increasing $x$ tends to decatalyze the condensate for larger 
$B$. The large $B$ behavior of the condensate is mainly determined by the backreaction of the flavor to the geometry. {\it Right: } The normalized chiral condensate as a function of $B$ in the confined phase ($T=0$) for different values of $x$. The condensate always increases, but for higher $x$ the rate of increase becomes smaller. At $x=0.01$ the background is essentially the same as the Yang-Mills case with no flavor. Then, the tachyon is affected only by the explicit $B$ dependence in Eq. (\protect\ref{tacheq}). As a result the condensate is catalyzed by $B$ for $x=0.01$.  }
\label{fig:qqConfSmallx} 
\end{figure}

Another clue comes from the fact that the implicit backreaction effect is controlled by the value of $x$. Indeed, as  B only enters the gravitational action in the flavor sector (\ref{actf}) and because this action is proportional to $x$, backreaction of B on the background functions should be bigger for larger values of $x$. Then, if our identification of this backreaction with the sea quark is correct, for a large value of B (where the backreaction effect is isolated) we should see suppression of  the condensate for larger values of $x$. In the left figure in \ref{fig:qqConfSmallx} we plot the dependence of the normalized condensate 
$\Delta \Sigma$ on $x$ for a large value of $B/\Lambda^2=3$. We indeed observe the suppression of the condensate with increasing $x$. 
In the right figure in \ref{fig:qqConfSmallx} we plot the dependence of $\Delta \Sigma$ on B for varying values of $x$. We find that the fortification of the condensate with B is much more pronounced for smaller $x$. This is again consistent with our identification, because the backreaction effect is absent for such small values and it should be the explicit dependence of the tachyon equation on B that controls the physics. It may seem confusing that the normalized condensate always increases with B in this figure. 
There is no contradiction however: as mentioned above, in the large $N_c$ limit the temperature dependence in the deconfined phase drops out, hence the plots in  \ref{fig:qqConfSmallx} are in fact for $T=0$. Indeed lattice studies also always find magnetic catalysis for small T. It would be nice to  study behavior in the deconfined and chirally broken phase at small $x$, but this phase does not exist in our model for small $x$, as can be seen from Fig. \ref{fig:varyingx}.

\section{Summary and Discussion}%
\label{sec:concl}

In this work we study the influence of an external magnetic field on a strongly interacting, confining theory of quarks and gluons in the large $N_c$, large $N_f$ limit with vanishing quark masses. We focused on two related problems: dependence of the quark condensate on the magnetic field, and the phase diagram of the theory on the B-T plane. We employed a bottom-up holographic model of QCD, known as V-QCD. This model perfectly suits our problem as it displays all the salient features of QCD: confines color and breaks the chiral symmetry at low temperatures, correctly reproduces the running of the coupling constant, exhibits a first order deconfinement and a second order chiral symmetry restoration transition with increased T and agrees almost perfectly with thermodynamic studies on the lattice. Holographic modeling necessitates the large $N_c$ limit. For finite number of flavors then, effect of the magnetic field on the system would be negligible. Thus we consider the Veneziano limit where the ratio $x= N_f/N_c$ is kept constant in the large-$N_c$ limit. The holographic model therefore includes the full backreaction of the flavor branes on the  background geometry. The model is necessarily complicated yet manageable. 

The magnetic field in our model is introduced by an Abelian bulk gauge field that corresponds to the diagonal U(1) of the flavor symmetry on the flavor branes. Coupling of this field to the background is controlled by two parameters: the ratio $x$ and a constant $c$ that parametrizes the gauge field kinetic term in the DBI action. The magnetic field influences the geometry stronger for larger values of $x$ and smaller values of $c$. We study the behavior of the quark condensate and the phase diagram of the theory for varying values of $c$ and $x$. Qualitative agreement with lattice studies follow for smaller choices of $c$. In particular, for smaller values of $c$, such as $c=0.4$, we observe the phenomenon of inverse magnetic catalysis in our model. We observe two manifestations of this phenomenon. First, the chiral and the deconfinement transition temperatures starts decreasing as the magnetic field is turned on. This is shown in figures \ref{fig:Tdc}, \ref{fig:conqq} and \ref{fig:varyingx}. Second, we observe that magnetic field decreases the value of the quark condensate in figures \ref{fig:conqq}, \ref{fig:Dqq}(left) for large enough T.

The phase diagram of the model is also interesting. As shown in figure \ref{fig:conqq} for $c=0.4$ and $x=1$ there are typically three phases in this model. At low temperatures the model is confined and the chiral symmetry is broken for all values of B. As we crank up the temperature we first come across a deconfinement phase transition, that is first order and the temperature it happens $T_d$ is a non-trivial function of B. As we crank T up further we hit a second order chiral symmetry restoration transition above which the condensate vanishes. This transition temperature $T_\chi$ also exhibits an interesting profile in B.  This picture becomes more complicated when we fiddle with the value of $x$. As we see from figure 
\ref{fig:varyingx} there arise the possibility of new phases for smaller value of $x$, such as $x=0.1$. This new phase shown in red is another deconfined and chiral symmetric phase separated from the usual one by a first order phase separation line. There also seems to be reappearance of the deconfined and chirally broken phase at larger values of B for $x=1/3$ and higher. Whether these features are realized in real physical systems or they are just artifacts of our holographic model remains to be seen. 

It is important to note that our model has the following limitation in the confined phase. Confined phase corresponds to the thermal gas geometry which does not exhibit any dependence on T in the large N limit. Therefore we cannot study dependence of the condensate on T in this phase and our plots are essentially for $T=0$. This is why we focus on the T-dependence in the deconfined and chiral symmetry broken phase, a phase that is generically present in our model. In order to track the T dependence also in the confined phase one needs to take into account the $1/N$ corrections, in particular the contribution to the free energy arising from the bulk fluctuations in the background. This is not an easy task that we hope to tackle in future. 

One of the main focus of our paper is to understand the physical mechanism behind inverse magnetic catalysis. We indeed find strong indications that this phenomenon is driven by the sea quark effect, that is the contribution of the quark determinant coming from the gluon path integral in $\langle \bar q q \rangle$. 
We identify this ``glue backreaction'' in the holographic model with the implicit dependence of the background functions on B that enter the tachyon equation. We find strong indications supporting this identification by analyzing the large B limit of the tachyon equation and varying the value of $x$, as discussed in the previous section. If this idea is indeed correct then the remaining question is why is this backreaction effect is causing the inverse effect? To answer this question we need to translate this holographic mechanism into the field theory language.  This is another interesting problem that we leave to future work. 

Shape of the phase separation curves at larger values of B is another open problem in the lattice and effective field theory studies. The lattice study \cite{Endrodi:2015oba} goes up to  $eB = 3$GeV and observes a monotonic decrease in the chiral transition temperature as a function of B up to these large values. General expectation from  effective field theory, however, is that eventually this separation line curves up and starts increasing with B \cite{Mueller:2015fka}. We generically observe this catalyzing behavior at larger values of B but the minimum of the phase separation line both depends on $x$ and $c$, hence it is model dependent. 

Finally we studied the various interesting observables such as magnetization and magnetic susceptibility. As shown in figure \ref{suscx1} there are interesting kinks appearing in the magnetic susceptibility as a function of T at the chiral transition. Comparison and perhaps matching the magnetic susceptibility in our model with lattice studies is also left as a future work. 

The most pressing issue is to fix the parametrization of our model, in particular fixing the shape of the potentials, shown in Appendix \ref{app:potentials}, by matching the lattice data. This is not a very easy task because ideally, we want our model to reproduce all the features observed on the lattice quantitatively. This requires matching at all different levels, including the meson spectra, the thermodynamic functions, transport coefficients and the various correlation functions. In particular, it will be also very interesting to also include a baryon chemical potential in our theory and fix the parameters of the model also including the finite baryon density domain. All of this we leave for future work.

\section{Acknowledgments} We thank Gunnar Bali, Pavel Buividovich, Falk Bruckmann,  Andreas Schafer and Ismael Zahed for useful discussions. This work was supported, in part by the Netherlands
Organisation for Scientific Research (NWO) under VIDI grant 680-47-518, and the Delta Institute
for Theoretical Physics (D-ITP) funded by the Dutch Ministry of Education,
Culture and Science (OCW).

\appendix

\section{The Equations of motion}\label{app:eoms}%

The Einstein equations of motion from the action (\ref{action}) read
\begin{align}
 R_{\m\n}-{1 \over 2} g_{\m\n} R - \left( {4\over 3} {\partial_{\m} \lambda \pa_{\n} \l \over \l^2} -{2 \over 3} {(\partial \l )^2 \over \l^2} g_{\m\n}  +{1\over 2} g_{\m\n} V_g(\l) \right)&& \nn\\
-x {V_f(\l,\t) \over 2} \left(- g_{\m\n} \sqrt{D} +{1 \over \sqrt{D}} {d D \over d g^{\m\n} }  \right)  &=0 \, , &
\end{align}
where $D=det(\delta_{\l}^{\m}+w(\l) \, g^{\m\n} \, V_{\n\l} + \kappa(\l)  g^{\m\n} \partial_{\n} \t \, \partial_{\l} \t )$.
Inserting here our Ans\"atze for the metric~\eqref{bame} and $V_\m$~\eqref{vans} gives 
\begin{align}
&3 A'' + {2\over 3} {\l'^2 \over \l^2} + 3 A'^2 + \left(3 A' -W'\right){ f' \over 2f}   +{x \, V_f(\l,\t) \, G \, e^{2 \, A} \over  2 \, Q \, f}(2 \, Q^2-1) -{ e^{2\, A} \over 2 f} V_g(\l)=0 \, , \nn \, \\
&W''+ {W' f'\over f} +W'^2 + 3 A' W'  + {x \, V_f(\l,\t) \,G \, e^{2 \, A} \over  2 \, Q\, f} \left( 1-Q^2\right)=0 \, , \, \label{backeq} \\
& f''+(3 A'+  W') \, f'+{x\, V_f(\l,\t) \, e^{2 A}\, G \over Q}\left(1-Q^2 \right)  =0 \nn \, ,
\end{align}
where we defined 
\begin{align}
 \label{Gdef}
 G(r) &= \sqrt{1 + e^{-2 A(r)}\kappa(\l,\tau) f(r) (\partial_r \t(r))^2} \,, \,\,\, \nn \\
   Q(r)&=\sqrt{1+w(\l)^2 B^2 e^{-4 A(r)}}\,.
\end{align}
The first order constraint equation reads
\begin{align}
&{2 \over 3}{\l'^2 \over \l^2}- \left(3 \, A' +W' \right) {f' \over 2 \, f}-6\,A'^2 -3\, A' \,W' +{ e^{2\, A} \over 2\, f} V_g(\l) -{x_f \, V_f(\l,\t) \, Q \, e^{2 \, A} \over 2 \, G \, f} =0 \,.
\end{align}
The dilaton equation of motion becomes
\begin{align}\label{laeq}
& {\l'' \over \l} -{\l'^2 \over \l^2}+\left(  3 A' +W' +   {f' \over f}   \right) {\l' \over \l} + {3 \over 8} {\l \, e^{2\, A} \over f} \, \partial_{\l} V_g(\l)
 - {3\, x \, B^2 \, e^{ -2 \, A}  \, G \, \l \, V_f(\l,\t) w(\l) \over 8 \, f \, Q}\partial_{\l}w(\l) \nn \\
&-{3 \, x \, e^{2 \, A}  \, G \, \l \, Q \over 8 f } \partial_{\l} V_f(\l,\t)
-{3 \, x \, \l \, Q \, V_f(\l,\t) \, \t'^2 \over 16 \, G } \partial_{\l} \kappa(\l)=0 \,,
\end{align}
and the tachyon equation of motion is
\begin{align} \label{tacheq}
&\t''-{e^{2 \, A} \, G^2 \over f\, \kappa(\l)}{\partial_{\t} \log \, V_f (\l,\t)}+\, e^{-2 \, A} \, f \, \k(\l) \, \bigg( W' +{1 \over 2} \,{ f' \over f}  \nn \\
&+ 2 A' {1+Q^2 \over Q^2} +{1 \over 2}\, \l' \, \partial_{\l} \log \,( \kappa(\l)  \,V_f(\l,\t)^2 ) - {\l' \,(1-Q^2)  \over Q^2} \partial_{\l} \, \log \, w(\l) \bigg) \t'^3 \nn \\
&+ \left(A'{2+Q^2 \over Q^2}+W' +{f' \over f} + \l' \partial_{\l} \log(V_f(\l,\t) \, \k(\l)) - {\l' \,(1-Q^2)  \over Q^2} \partial_{\l} \, \log \, w(\l) \right) \,\t' =0 \,. \nn \\
\end{align}
The equation of motion of the vector gauge field is trivially satisfied for the Ansatz (\ref{vans}). The above equations of motion enjoy the following scaling symmetries

\begin{itemize} \label{rescaling}
\item rescaling of $f(r)$, 
$f \to {f \over c_f} \, ,\,\, A \to A-{1\over 2}\log \,c_f \, , \,\, B\to{B \over c_f}$
\item  rescaling of the holographic coordinate, 
$r \to \Lambda \, r \,\, , \,\,\, A \to A - \log \, \Lambda \,, \,\, B\to {B \over \Lambda^2}$
\item rescaling of the function $W(r)$, 
$W \to W +  c_{W}$ \,.
\end{itemize}

\section{The Potentials}\label{app:potentials}%
The potentials entering the action of the holographic model are the $V_g(\lambda)$, $V_f(\lambda, \tau)$, $w(\lambda)$, and $\kappa(\lambda)$. The dilaton potential $V_g(\lambda)$ governs the glue dynamics in the absence of flavors, the tachyon potential, $V_f(\lambda, \tau)$  is mainly responsible for the dynamics of the tachyon condensation and the breaking of chiral symmetry at zero temperature. Additionally, the potentials $w(\lambda)$ and $\kappa(\lambda)$ determine the coupling of the mesons to glue. The choice of the potentials in the current 
work coincides with that of \cite{altemu}. The near boundary expansion of those potentials is such that the perturbative dynamics of QCD are reproduced. In particular, the numerical coefficients of $V_g, V_{f0}$ and $\kappa$ are fixed by matching to the beta function of QCD and the quark mass anomalous dimension.  The explicit form of the potentials is

\begin{eqnarray}
\label{Vf0SB}
V_g(\lambda)&=&{12\over \mathcal{L}_0^2}\biggl[1+{88\lambda\over27}+{4619\lambda^2
\over 729}{\sqrt{1+\ln(1+\lambda)}\over(1+\lambda)^{2/3}}\biggr]\, , \\
 V_{f0}& =& {12\over \mathcal{L}_{UV}^2}\biggl[{\mathcal{L}_{UV}^2\over\mathcal{L}_0^2}
-1+{8\over27}\biggl(11{\mathcal{L}_{UV}^2\over\mathcal{L}_0^2}-11+2x \biggr)\lambda\nn\\
 &&+{1\over729}\biggl(4619{\mathcal{L}_{UV}^2\over \mathcal{L}_0^2}-4619+1714x - 92x^2\biggr)\lambda^2\biggr] \, , \nn \\
 \kappa(\l) &=& {[1+\ln(1+\l)]^{-1/2}\over[1+\frac{3}{4}(\frac{115-16x }{27}-{1\over 2})\l]^{4/3}} \,, \quad\quad a(\l)=\frac{3}{2 \, \mathcal{L}_{UV}^2} \, ,
\label{kappaa}
 \end{eqnarray}
where $\mathcal{L}_{UV}$ is  the AdS radius, so that the boundary expansion of the metric is 
$ A \sim \ln \left( { \mathcal{L}_{UV} / r} \right)+\cdots \, .$
The radius depends on $x$ as 
\be
\mathcal{L}_{UV}^3 = \mathcal{L}_0^3 \left( 1+{7 x \over 4} \right) \, .
\label{adsrad}
\ee
The IR asymptotics of the potentials are constrained by certain low energy features of the dual field theory. Those include confinement, chiral symmetry breaking and the correct thermodynamic behavior of the theory at strong coupling at finite $N_f/N_c$. Moreover, the behavior of the glueball and meson spectra is highly dependent on the IR behavior.  
For the function $w$ we use the choice
\be
w(\l)=\kappa(c\l) =  \frac{( 1+\log(1+ c \, \l))^{-{1\over 2}}}{\left(1+ {3 \over 4} \left({115-16 x \over 27}-{1\over 2} \right) c \,\l  \right)^{4/3}}       \, .
\label{wl}
\ee

\bibliographystyle{JHEP}
\bibliography{Bref}

\end{document}